\documentclass[prb,12pt]{revtex4}

\usepackage[english]{babel}
\usepackage[intlimits]{amsmath}
\usepackage[dvips]{graphicx}

\begin{document}

\begin{titlepage}

\title{Anisotropic Lattice Models of Electrolytes}
\author{ Vladimir Kobelev and  Anatoly B. Kolomeisky}
\affiliation{Department of Chemistry, Rice University, Houston, TX 77005}

\begin{abstract}

\noindent Systems of charged particles on anisotropic three-dimensional lattices  are investigated theoretically using  Debye-H\"uckel theory. It is found that the thermodynamics of these  systems strongly depends on the degree of anisotropy. For weakly anisotropic simple cubic lattices,  the results indicate the existence of order-disorder phase transitions and  a tricritical point, while the possibility of  low-density gas-liquid coexistence is suppressed. For strongly  anisotropic lattices  this picture changes dramatically: the low-density gas-liquid phase separation reappears and the  phase diagram exhibits  critical, tricritical and triple points. For body-centered lattices, the low-density gas-liquid phase coexistence is suppressed for  all degrees of anisotropy. These results show that the effect of anisotropy in lattice models of electrolytes amounts to reduction of spatial dimensionality.
\end{abstract}

\maketitle

\end{titlepage}

\section{Introduction}

Understanding  thermodynamic properties of electrolyte systems is a long-standing problem\cite{debye,fisher94,weingartner01} which, in recent years, has attracted  increased attention  due to controversial results on the nature of criticality in Coulomb systems.\cite{fisher94,weingartner01} Experiments\cite{pitzer} suggest that the critical region of solutions of some organic salts  can be well described by classical behavior, while for other electrolyte systems the Ising-like description with non-classical behavior  is more appropriate.\cite{japas} This  dichotomy has greatly stimulated theoretical attempts to understand the thermodynamics of ionic systems.

Criticality in simple nonionic fluids can be successfully described and analyzed by the renormalization group (RG) method, and it is reasonable to suggest that this method could also be used  to investigate also  Coulomb  systems.  However, in order to proceed with RG calculations, a physically meaningful and well based mean-field theory should be developed.\cite{fisher94} Most  theoretical studies of ionic systems concentrate on the simplest model of electrolytes, the so called restricted primitive model (RPM), in which ions are viewed as equal size particles of positive and negative charges of equal magnitude. Currently, there are two theoretical directions in the  development of the mean-field description of ionic fluids.  The first approach is based on integral equations for correlation functions\cite{stell95,yeh96,ciach01,blum02,brognara02}, while the second approach extends  the original Debye-H\"{u}ckel (DH) theory.\cite{fisher94,levin94,levin96,zuckerman97,kobelev02} Comprehensive theoretical analysis,\cite{levin94,levin96,zuckerman97}  which utilizes for example thermodynamic energy bounds,  and comparison with current Monte Carlo simulations,\cite{pana02,caillol02}  indicate that theories based on DH theory  may provide a better description of the thermodynamics of electrolytes in critical regions.   

So far most of the theoretical efforts in the investigation of charged systems have been  devoted to continuum  models. However, lattice models are also important for understanding the criticality in Coulomb systems, since the Ising model, which is a lattice gas model, has been crucial  for the  description of  critical phenomena in nonionic systems.\cite{fisher94,weingartner01}  There are few numerical\cite{pana99,dickman99} and analytical\cite{ciach01,brognara02,kobelev02} results for  lattice ionic systems  which show that the phase diagram differs significantly from continuum models. The  structures of simple cubic (sc) and body-centered cubic (bcc) lattices allow for charge distribution with the appearance of a long-range order phase at low temperatures similar to that of an ionic crystal. However, this ordering decreases entropy and for high  temperatures the disordered phase is  thermodynamically more stable. As a result,  there is an order-disorder phase transition line  which ends up at the tricritical point,\cite{pana99,dickman99,brognara02,kobelev02} while for continuum RPM systems only a  gas-liquid coexistence can be found. 

Recently, a systematic investigation of lattice models of electrolytes has been presented.\cite{kobelev02} In this investigation the lattice restricted primitive model (LRPM), with charged particles occupying sites on a general $d$-dimensional lattice, has been considered using Debye-H\"{u}ckel theory.  By solving exactly the lattice version of the  Debye-H\"{u}ckel equation, closed  expressions for thermodynamic properties of general $d$-dimensional ionic systems have been obtained. For three-dimensional lattice Coulombic systems specific calculations, which included pairing and dipole-ion solvation,  yielded a gas-liquid phase separation at low densities. However, by taking into account the lattice symmetry, it has been shown that for sc and bcc lattices this gas-liquid phase separation is thermodynamically unfavorable, and  the order-disorder phase transitions with the tricritical point will dominate, in agreement with Monte Carlo simulation results.\cite{pana99,dickman99}

The thermodynamics of lattice anisotropic  ionic systems have not been studied yet, although  they may provide  important information on thermodynamics of real electrolytes. In addition, the lattice  stretching, which leads to anisotropy, can be viewed as analogous of lowering the spatial dimensionality of the system. However, DH-based calculations for the continuum\cite{levin94} and lattice\cite{kobelev02} models of electrolytes predict an {\it increase} of gas-liquid critical temperatures for lower dimensions. Thus, for lattice anisotropic models,  gas-liquid phase separation may reappear along with distinct order-disorder phase transitions. This possibility raises the question of the precise determination of  phase diagrams for  ionic systems  on anisotropic  lattices.

In this article, we present a theoretical investigation of anisotropic lattice models of electrolytes using the Debye-H\"{u}ckel method,\cite{kobelev02}  which has the  advantage that it accounts for both electrostatic screening and sublattice ordering in a unified framework. We investigate the ionic systems on  three-dimensional lattices obtained by anisotropic stretching of simple cubic and body-centered cubic lattices. The paper is organized as follows. In Sec. II we provide  a combined DH and mean-field ordering description of ions  on  simple tetragonal  lattices. The similar analysis for  stretched body-centered lattices is given in Sec. III. A discussion and conclusions  are presented in Secs. IV and V.

\section{Debye-H\"uckel Theory for  Simple Tetragonal Lattices} 

\subsection{Pure DH theory}

Our derivation for anisotropic lattice electrolytes follows closely the  Debye-H\"uckel approach for isotropic lattice ionic systems.\cite{kobelev02} We consider a system of equal numbers of positive and negative ions with the total density $\rho=\rho_-+\rho_+$ on a tetragonal lattice, which is obtained by stretching the simple cubic lattice,  with unit cell dimensions $a\times a\times b$ with an anisotropy factor defined as the  ratio of lattice parameters, $\alpha=a/b$. The case $\alpha=1$ corresponds to the isotropic simple cubic lattice electrolytes which have been studied in detail earlier.\cite{kobelev02} The  Debye-H\"uckel approach implies that we can construct the total  free energy of the system by summing consecutively the terms which describe  interactions between different species. Charged particles interact through the lattice Coulomb potential, otherwise they behave  as ideal particles with additional hard-core on-site exclusions. Thus the total free energy density  is given by  $f=f^{Id}+f^{DH}$.  The ideal lattice gas  contribution  can be written as 
\begin{equation} \label{f.id}
    \bar f^{Id} = -\frac{F}{k_{B}TV}=-\frac{\rho^{*}}{v_{0}} \ln \rho^* -\frac{(1-\rho^*)}{v_{0}} \ln(1-\rho^*),
\end{equation}
where $\rho^*=\rho v_0$ is the reduced dimensionless density and $v_0=a^2b$ is the  unit lattice cell volume.

The other term in $f$  comes from the Coulombic interactions of the free ions and  includes the effect of  screening. In order to find an expression for the Debye-H\"uckel contribution to the free energy density, we need  the  electric potential felt by an ion due to all other ions. This potential  can be found by fixing an arbitrary ion at the origin and solving the linearized Poisson-Boltzmann equation 
\begin{equation} \label{linPB}
   \Delta \varphi(\mathbf{r})=\kappa^2 \varphi(\mathbf{r})-(C_d q /D v_0)\delta(r),  
\end{equation}
where $\kappa^2 = C_d \beta \rho q^2 /D $ is the inverse squared Debye screening length, with $\beta=1/k_{B}T$ and $C_d=4\pi$ for three-dimensional lattices. 
The lattice Laplacian used in Eq.(\ref{linPB}) can be presented  in the form which incorporates the geometry of the lattice, 
\begin{equation} \label{laplac1}
    \Delta \varphi= \Delta_x \varphi+\Delta_y \varphi+\Delta_z \varphi,
\end{equation}
with
\begin{equation} \label{laplac2}
   \Delta_i \varphi(\mathbf{r}) = 1/a_i^2 \left[\varphi(\mathbf{r}-a_i\mathbf{e}_i)-2\varphi(\mathbf{r})+\varphi(\mathbf{r}+a_i\mathbf{e}_i)\right],  
\end{equation}
where $i=x,y,z$; $a_x=a_y=a,  \ a_z=b$ and $\mathbf{e}_i$ are the unit vectors along the corresponding lattice directions. Then Eq.(\ref{linPB}) can be easily solved by Fourier transformation to yield for $r>0$
\begin{equation} \label{fiofr}
   \varphi(\mathbf{r})=\frac{2\pi q a^2}{3 D v_0} \int_k \frac{e^{i \mathbf{k r}}}{(x^2+4+2\alpha^2)/6- J(\mathbf{k},\alpha^2)},
\end{equation}
where we introduced the anisotropic lattice function
\begin{equation} \label{j.an}
   J(\mathbf{k},\gamma)= \frac{1}{c_{0}} \sum_{nn} e^{i\mathbf{k} \cdot \mathbf{a}} = \frac{1}{3} (\cos k_1 + \cos k_2 + \gamma \cos k_3),
\end{equation}
$c_{0}=6$ is the number of nearest neighbors, and $\int_k \equiv (2\pi)^{-d}\int_{-\pi}^{\pi}d^d \mathbf{k}$ and $x=\kappa a$. Note  that the vectors $\mathbf{k}=(k_{1}/a, k_{2}/a, k_{3}/b)$ describe the reciprocal lattice.

To compute the electric potential at the origin due to the surrounding ions, it is recalled  that no other ion can be placed at that site, and hence the appropriate equation is the Laplace's equation,
\begin{equation} \label{eq.lap}
    \Delta \varphi(\mathbf{r}=0)=0.
\end{equation}
This enables us to write
\begin{equation}
  \Delta_x \varphi(\mathbf{r}=0)=\Delta_y \varphi(\mathbf{r}=0)=\Delta_z \varphi(\mathbf{r}=0)=0,
\end{equation}
which yields
\begin{eqnarray} \label{fiof0.1}
    \varphi(\mathbf{0})&=&(\varphi(\mathbf{r}-a\mathbf{e}_x)+\varphi(\mathbf{r}+a\mathbf{e}_x))/2 \nonumber\\
                       &=&(\varphi(\mathbf{r}-a\mathbf{e}_y)+\varphi(\mathbf{r}+a\mathbf{e}_y))/2 \\
                       &=&(\varphi(\mathbf{r}-a\mathbf{e}_z)+\varphi(\mathbf{r}+a\mathbf{e}_z))/2 \nonumber
\end{eqnarray} 
Then from Eqs. (\ref{fiofr}) and (\ref{fiof0.1}) it follows that
\begin{equation} \label{fiof0.2}
   \varphi(\mathbf{0})=\frac{2\pi q a^2}{D v_0 (2+\alpha^2)} \int_k \frac{J(\mathbf{k},\alpha^2)}{(x^2+4+2\alpha^2)/6- J(\mathbf{k},\alpha^2)}.
\end{equation}
Following the DH approach for isotropic lattices,\cite{kobelev02} we introduce  the integrated  anisotropic lattice Green's function
\begin{equation} \label{green.an}
  P(z,\alpha)= \int_{\mathbf{k}} \frac{1}{1-z J(\mathbf{k}, \alpha^2)},
\end{equation} 
which has been  evaluated exactly in terms of a product of two complete elliptic integrals of the first kind  by Joyce.\cite{joyce01}  Then the total potential at the origin due to the surrounding ions, $\psi(0)=\varphi(0)-\varphi(0)|_{x=0}$, takes the form
\begin{equation} \label{psi.1}
    \psi = \frac{2 \pi q}{D b (2+\alpha^2)}\left[P\left(\frac{6}{x^2+4+2\alpha^2},\alpha\right)-P\left(\frac{6}{4+2\alpha^2},\alpha\right)\right].
\end{equation}

By using the  Debye charging procedure, the reduced electrostatic free energy density can be calculated explicitly, yielding  
\begin{equation} \label{f.dh}
   \bar f^{DH}=\frac{1}{4(2+\alpha^2) v_0}\left[ x^2 P\left(\frac{6}{4+\alpha^2},\alpha\right) - \int_0^{x^2}P\left(\frac{6}{x^2+4+\alpha^2},\alpha\right) d(x^2) \right].
\end{equation}
The  chemical potential, $\bar{\mu}=\mu/k_{B}T= -\partial \bar{f}/\partial \rho$, is then given by
\begin{equation} \label{chem.potential}
\bar \mu =  \ln \rho^{*} -\ln(1-\rho^{*}) - \frac{\pi}{(2+\alpha^2)T^{*}}\left[  P\left(\frac{6}{4+\alpha^2},\alpha\right) - P\left(\frac{6}{x^2+4+\alpha^2},\alpha\right)  \right],
\end{equation} 
where, following the continuum DH theory\cite{levin96} and the DH theory for isotropic cubic lattices,\cite{kobelev02}  the reduced temperature is defined as
\begin{equation}  \label{temp}
   T^* = \frac{D k_B T v_0}{q^2 a^2} = \frac{D b}{q^2 \beta}=\frac{D a}{q^2 \beta \alpha},
\end{equation}
and for the reduced density we  obtain
\begin{equation} \label{density}
   \rho^*=\frac{x^2 T^{*}}{4 \pi}.
\end{equation}
Knowing the free energy density and the chemical potential allows us to calculate the pressure, $\bar p = p/k_{B}T= \max_{\rho}[\bar{f}+\bar{\mu}\rho]$, yielding 
\begin{equation} \label{pressure}
    \bar pv_{0}= -\ln(1-\rho^*) + \frac{1}{4 (2+\alpha^2) } \left[x^2 P\left(\frac{6}{x^2+4+2\alpha^2},\alpha\right) - \int_0^{x^2}P\left(\frac{6}{x^2 +4+2\alpha^2},\alpha\right)d(x^2)\right]. 
\end{equation}

Eqs. (\ref{f.id}), (\ref{f.dh}), (\ref{chem.potential}) and (\ref{pressure}) provide a full thermodynamic description of the simple tetragonal lattice model of electrolytes. The thermodynamics at the critical region can be investigated by analyzing the spinodal, which is determined by the condition $\rho \frac{\partial \bar{\mu}}{\partial \rho}=0$. Using Eq. (\ref{chem.potential}) we obtain
\begin{equation} \label{spino}
     T^*_{s} =\frac{2\pi}{(2+\alpha^2)}\; \frac{\zeta(1-\zeta) \partial P(\zeta)/\partial \zeta}{2+(1-\zeta)^2 \partial P(\zeta)/\partial \zeta}, 
\end{equation}
with $\zeta=6/(x^2+4+2\alpha^2)$.

The phase transitions and the gas-liquid coexistence can be studied by analyzing the pressure and the chemical potential in different phases. The predicted  gas-liquid coexistence curves for  simple tetragonal lattices  are shown  in Fig.1. The critical temperature  increases monotonically as the anisotropy parameter decreases and  reaches the value of  $T_{c}^{*}=1/2$ at $\alpha=0$: see Fig.2a. At the same time, the critical density shows a non-monotonic behavior with two  maxima and a minimum, and finally approaches the value   $\rho_{c}^{*}=0$ at  $\alpha=0$, as shown in  Fig.2b. Lowering the anisotropy parameter can be visualized as  the stretching  the lattice along one direction, and the limit of $\alpha \rightarrow 0$ corresponds to an  infinite distance between the layers. Thus the anisotropic lattice model at $\alpha=0$ is equivalent to the 2D Coulomb system on square  lattice, for which the pure DH theory\cite{kobelev02} predicts the critical parameters to be $T_{c}^{*}=1/4$ and $\rho_{c}^{*}=0$. The apparent discrepancy between   our results for the critical temperature and the results for two-dimensional lattice electrolytes\cite{kobelev02} can be easily explained by analyzing the linearized Poisson-Boltzmann Eq.(\ref{linPB}). In our calculations we used the three-dimensional coefficient $C_{d}=4 \pi$, while in two dimensions this coefficient  is equal to $2\pi$, which explains the factor 2 in the difference in the corresponding values of the critical temperatures. 

At the limit of large $\alpha$ the lattice is stretched along 2 directions, and thus the anisotropic lattice with $\alpha = \infty$ corresponds to the  1D lattice Coulomb system. Our predictions for critical parameters in this case are $T_{c}^{*}=0$ and $\rho_{c}^{*}=0$, while for one-dimensional lattice electrolytes the DH-based  calculations give  $T_{c}^{*}= \infty$ and $\rho_{c}^{*}=0$. In this case, the difference in critical temperatures can be attributed to our definition of the reduced temperature in Eq.(\ref{temp}). Note, however, that the DH method is incorrect in describing one-dimensional ionic systems.\cite{kobelev02} The overall agreement between our estimates of critical parameters of strongly anisotropic lattice models of electrolytes and the results for 1D and 2D ionic lattice systems supports our arguments that anisotropic stretching is analogous to lowering of spatial dimensionality for simple cubic lattices. 

A surprising feature of  the predicted coexistence curves is the critical density dependence on the  lattice anisotropy as exhibited in Fig.2b. It  shows two maxima (at $\alpha=0.4$ and $\alpha=4.175$) and one minimum (at $\alpha=1$);  this picture is probably   the result of geometric and packing effects.

\subsection{Sublattice Ordering}

The Debye-H\"uckel  approach presented above describes only the low-density behavior of the system. To obtain the full thermodynamic description of  simple tetragonal lattices we have to take into account the lattice symmetry. A  simple tetragonal  lattice, similarly to a simple cubic lattice,  can be viewed as consisting of two intercalated sublattices. Ions of opposite signs can be distributed unequally between these sublattices, thus   reducing the electrostatic contribution to the free energy. At the same time, unequal distribution of charged particles   lowers  the entropy which leads to an increase in  the  total free energy. The competition between these factors determine the thermodynamics and phase behavior of the system.

We consider again a simple tetragonal  cubic  lattice with $N$ charged particles. The overall system is neutral, and there are $N_{A}^{+}$ ($N_{A}^{-}$) positive (negative) particles in sublattice $A$, and   $N_{B}^{+}$ ($N_{B}^{-}$) positive (negative) particles in sublattice $B$. Assuming that sublattice $A$ has an excess of positive ions, the corresponding order parameter can be defined as
\begin{equation}
  y=\frac{N^+_A-N^-_A}{N^+_A+N^-_A}=-\frac{N^+_B-N^-_B}{N^+_B+N^-_B}.
\end{equation}
This order parameter  has a  positive value  in the ordered phase, while it equals to zero in the disordered phase. 

The nonzero charge density on each sublattice produces an additional ``background'' potential $\Phi(\mathbf{r})$ which, however, does not change the correlation functions.\cite{kobelev02} This potential can be found using the linearized Poisson-Boltzmann equation 
\begin{equation}
   \Delta \Phi(\mathbf{r}_A)= -(4\pi/D) \rho y q .
\end{equation}
Because of the symmetry between  sublattices, we have  $\Phi(\mathbf{r}_A)=-\Phi(\mathbf{r}_B)$. By using the definition of the lattice Laplacian (\ref{laplac1})-(\ref{laplac2}), and following the approach outlined for isotropic lattice electrolytes,\cite{kobelev02}  we obtain for the potential $\psi$ due to all ions except the one fixed at the origin the following expression
\begin{equation}
  \psi(\mathbf{r}_A)=-\frac{\pi}{2+\alpha^2} \frac{\rho^* y q}{D b} +\psi^{DH},
\end{equation}
where $\psi^{DH}$ is given in Eq.(\ref{psi.1}).  The electrostatic part of the total  free energy then follows again from the Debye charging process, while the entropic contribution can be calculated as for isotropic lattices,\cite{dickman99,kobelev02} yielding $ \bar f = \bar f^{Id}+ \bar f^{DH} + \bar f^{Ord}$ with
\begin{equation} \label{f.ord}
   \bar f^{Ord} = \frac{\pi}{2(2+\alpha^2)}\frac{\rho^* y^2}{v_0 T^*} - \frac{\rho^*}{2 v_0}\left[(1+y)\ln(1+y)+(1-y)\ln(1-y)-2\ln2\right],
\end{equation}
where $\bar f^{Id}$ and $\bar f^{DH}$ are given by Eqs.(\ref{f.id}) and (\ref{f.dh}).

The knowledge of the total free energy allows us to investigate the possibility of sublattice ordering. It can be done by looking for minima of $\bar f^{Ord}$ for nonzero values of the order parameter $y$. This procedure  leads us to the equation describing the $\lambda$-line, along which  second-order phase transitions occur,
\begin{equation}
   \rho_\lambda^*=\frac{2+\alpha^2}{\pi}T^*.
\end{equation}
The anticipated tricritical point can be found by calculating  the  intersection of the  $\lambda$-line with the spinodal $\partial \bar p/\partial \rho^*=0$, and this analysis  yields the equation for the tricritical point,\cite{dickman99,kobelev02}
\begin{equation} \label{tri}
    \frac{4(2+\alpha^2)}{\rho_{tri}^*}\left.\left[\frac{\partial P_s[6/(x^2+4+2\alpha^2),\alpha]}{\partial (x^2)}\right]\right|_{x^2=4(2+\alpha^2)}+\frac{1}{1-\rho_{tri}^*} - \frac{3}{2}=0.
\end{equation}

The resulting tricritical densities and temperatures for different values of the anisotropic parameter $\alpha$ are presented in Figs.2a and 2c. The tricritical temperature is a decreasing function of the anisotropy parameter, and it  reaches its maximal value of $T_{tri}=0.5960$ at $\alpha=0$, which is exactly twice the value of the tricritical temperature for the ionic system on the  two-dimensional  square lattice (see Eq.(70) of Ref. \cite{kobelev02}). This deviation is again the result of using different dimension-dependent coefficients $C_{d}$, as was argued above. Both critical and tricritical temperatures  vanish at large anisotropies, however, $T_{tri}$ becomes smaller than $T_{c}$ for  $\alpha >4.25$. The behavior of the tricritical density is different. It has a minimal value for the isotropic  lattice ($\alpha=1$), and  it reaches the maximal values of $\rho_{c}=0.3794$ and $\rho_{c}=0.416$ for $\alpha=0$ and $\alpha=\infty$ respectively. 

The phase diagrams for simple tetragonal lattices are presented in Fig.3. For weakly anisotropic lattices there are only order-disorder phase transitions, while for strongly anisotropic lattices  the gas-liquid phase separation reappears at low densities.

\section{Theory for  Body-Centered Tetragonal Lattice}

Consider a system of equal numbers of positive and negative ions on the  body-centered tetragonal  lattice with $2a \times 2a \times 2b$ unit lattice cell. Using the symmetry of the lattice and applying Eqs.(\ref{laplac1}) and (\ref{laplac2}) the lattice Laplacian for tetragonal body-centered  lattice is given by
\begin{equation} \label{bc.laplac}
  \Delta \varphi(\mathbf{r}) = \frac{2 (2+\alpha^2)}{3 c_0 a^2}\sum_{\mathbf{a}_{nn}}\left[\varphi(\mathbf{r}+\mathbf{a}_{nn})-\varphi(\mathbf{r})\right],
\end{equation}
where $\alpha=a/b$ and the summation runs over all $c_0=8$ neighbors in a cell. When $a=b$ this reduces to the well known lattice Laplacian for the body centered cubic lattice.\cite{katsura71,kobelev02} Because of the special symmetry of body-centered  lattice, the lattice function  
\begin{equation} \label{lattice.bcc.function}
J_b(\mathbf{k})=\frac{1}{c_{0}} \sum_{nn} e^{i \mathbf{k} \cdot \mathbf{a}}=\cos k_1 \cos k_2 \cos k_3,
\end{equation} 
in contrast to the simple tetragonal  lattice, is independent of the anisotropy parameter $\alpha$. Thus the isotropic bcc lattice Green's function $P_b(z)=\int_\mathbf{k}\frac{1}{1 - z J_b(\mathbf{k})}$ can be used for calculation of the thermodynamic properties. Then the potential at the origin due to the surrounding ions takes the form
\begin{equation} \label{pot.b}
    \psi=\frac{2 \pi q}{D v_0}\frac{3 a^2}{(2+\alpha^2)} \left[P_b\left(\frac{4+2\alpha^2}{x^2+4+2\alpha^2}\right)-P_b(1) \right],
\end{equation}
and the electrostatic free energy  is given by
\begin{equation} \label{f.dh.b}
   \bar{f}^{DH}= \frac{1}{4(2+\alpha^2) v_0} \left[x^2 P_b(1)-\int_0^{x^2}P_b\left(\frac{4+2\alpha^2}{x^2+4+2\alpha^2}\right)d(x^2)\right].
\end{equation}
Furthermore, the ordering free energy does not depend on the type of lattice \cite{kobelev02} and  is the same both for the simple and body centered lattices. Therefore we can use the corresponding expression from (\ref{f.ord}) and estimate the tricritical point from Eq.(\ref{tri}) by using $P_b(\frac{4+2\alpha^2}{x^2+4+2\alpha^2})$ instead of the simple tetragonal Green's function $P[6/(x^2+4+2\alpha^2),\alpha]$. Phase diagrams for electrolytes on the tetragonal body-centered  lattices are shown in Fig.4.

Analysis of phase diagrams for the  body-centered  lattice models of ionic systems indicate that both critical and tricritical densities are independent of the degree of lattice  stretching, while $T_{c}^{*}$ and $T_{tri}^{*}$ are decreasing functions of $\alpha$. These relations  can be understood by analyzing the corresponding equation for the  spinodal,
\begin{equation} \label{spino.bcc}
     T^*_{s} =\frac{2\pi}{(2+\alpha^2)}\; \frac{\zeta(1-\zeta) \partial P_b(\zeta)/\partial \zeta}{2+(1-\zeta)^2 \partial P_b(\zeta)/\partial \zeta}, 
\end{equation}
where $\zeta=(4+2 \alpha^2)/(x^2+4+2\alpha^2)$. Since the lattice Green's function $P_b(\zeta)$ is independent of the anisotropy parameter $\alpha$, the critical point can be described by  the parameter $x_{c}^2=2(2+\alpha^2)(1-\zeta_{c})/\zeta_{c}$ with $\zeta_{c}$ being {\it independent} of $\alpha$, while $T_{c}^{*} \propto 1/(2+\alpha^2)$. Then, utilizing Eq.(\ref{density}) the critical density $\rho_{c}=x_{c}^{2} T_{c}^{*}/4\pi$ is also independent of $\alpha$. Similar arguments can now  be applied for the analysis of the tricritical point. The thermodynamic behavior of the tetragonal body-centered lattice electrolytes is then different from that of the  simple tetragonal lattice ionic systems. Since  critical and tricritical temperature decay at the same rate  as  functions of the anisotropy parameter $\alpha$, while $\rho_{c}^{*}$ and $\rho_{tri}^{*}$ are constant, the possible low-density gas-liquid coexistence is suppressed by order-disorder phase transitions with a tricritical point at any degree of anisotropy, as shown in Fig.4.

\section{Discussion.}

Our analysis of  simple tetragonal lattice models of electrolytes based on the DH approach indicates that, similarly to isotropic sc ionic lattice systems,  at weak anisotropies ($0.385<\alpha<2.113$) the possible low-density gas-liquid phase separation is  metastable, and the sublattice ordering is  always thermodynamically more favorable. However, for strongly anisotropic simple  lattices ($\alpha<0.385$ or $\alpha>2.113$) the gas-liquid coexistence reappears and the phase diagram becomes more complex,  with critical, tricritical and triple points, as shown in Fig.3. The explanation of this phenomenon is the following. For weakly anisotropic lattices  the ordering of ions of opposite signs on different sublattices decreases significantly the total free energy, and order-disorder phase transitions with a tricritical point determines  the phase diagram of the system. For strongly anisotropic lattices this ordering is less significant at low densities and the gas-liquid phase separation is restored. Another way of  looking at this phenomenon, as was discussed above, is the analogy between lattice stretching and lowering of the space dimensionality. As was shown before,\cite{kobelev02} at low dimensions the critical temperature is increasing and thus   gas-liquid coexistence occurs  again  at low densities.
 
However, the thermodynamic behavior of Coulomb systems on  body-centered tetragonal lattices is very different. At all degrees of anisotropy, the sublattice ordering is  thermodynamically more stable, and  gas-liquid phase separation is always suppressed. This is the result of the special symmetry of body-centered  lattices.

Our pure  Debye-H\"{u}ckel treatment of anisotropic lattice electrolytes assumed that there are only free ions and empty sites in the system. However, at low temperatures the formation of strongly bound neutral dimers, or Bjerrum pairs,\cite{bjerrum,levin96,kobelev02} is a highly favorable process. Such pairing  can be viewed as a reversible chemical reaction,\cite{fisher94,levin96,kobelev02} and this process  can be treated in a systematic way. Another important contribution to the free energy of electrolytes is the ion-dipole solvation energy.\cite{fisher94,levin96,kobelev02}  The formation of ion pairs and their interactions with single ions  have a strong effect on the thermodynamic properties and phase diagrams of electrolytes.\cite{levin96,kobelev02} Our theoretical method can be extended to include these effects, although, due to anisotropy of the lattice system, there will be more than one type of bound neutral dimers. Based on the comparison with the continuum and isotropic lattice electrolytes,\cite{levin96,kobelev02} we predict that, if we take these  effects into account, both critical and tricritical temperatures will decrease, while  critical and tricritical densities will increase.  However, this means that our qualitative conclusions on thermodynamics and phase diagrams of anisotropic cubic lattices will not change.

It is interesting to note that similar phase diagrams of lattice electrolytes have been obtained by Ciach and Stell.\cite{ciach01} They considered a mean-field theory of electrolytes with single-ion lattice potential   on the  isotropic sc lattice and  additional short-range interactions. Note  that this  treatment neglects the cooperative screening, which is thermodynamically important and which is included in our DH-based theory. However, the origins  of similar complex phase diagrams in both models are different. In the model of  Ciach and Stell  the gas-liquid phase coexistence is driven by short-range interactions, while in our model the lattice stretching makes the sublattice ordering less thermodynamically favorable at low densities and,  as a result,  gas-liquid phase separation is restored.

\section{Conclusions}

We have extended the  Debye-H\"uckel method  to treat three-dimensional anisotropic    lattice models of electrolytes. Phase diagrams for different degrees of anisotropy have been obtained. For weakly anisotropic simple tetragonal lattices, the order-disorder phase transitions with a  tricritical point suppress the possibility of low-density gas-liquid phase transitions. However, for strongly anisotropic lattices  gas-liquid phase coexistence is restored. Thus the lattice anisotropy for simple tetragonal lattices  mimics  the lowering of the  spatial dimensionality. However, the thermodynamics of  the body-centered tetragonal  lattice ionic systems is very different. There is no gas-liquid separation and the phase diagram has only order-disorder phase transitions for  all degrees of anisotropy. This is the consequence of the special symmetry of  body-centered  lattices.

The relevance of  our results for understanding  the thermodynamics of real ionic fluids remains unclear. However, our method may be more useful for description of the thermodynamics of real ionic crystals with defects.\cite{crystal}  Furthermore, numerical simulations of the  anisotropic lattice ionic  models with lattice Coulombic potentials are clearly needed in order to check the validity of our theoretical predictions.

\section*{ACKNOWLEDGMENTS}

Acknowledgement is made to the Donors of the American Chemical Society Petroleum Research Fund (Grant \#37867-G6) for support of this research. We  also acknowledge  the support of the Camille and Henry Dreyfus New Faculty Awards Program (under Grant No. NF-00-056). We would like to thank Prof. M.E. Fisher, Prof. A.Z. Panagiotopoulos and Prof. M. Robert  for critical comments and useful suggestions.

\newpage

\noindent{\bf Figure Captions}\\
\\

\noindent Fig.1.  Gas-liquid coexistence curves for  simple tetragonal  cubic  lattices predicted by pure DH theory for different values of the lattice anisotropy parameter  $\alpha$.\\

\noindent Fig.2. Critical parameters as a function of degree of anisotropy: (a) critical  and tricritical  temperatures; (b) critical density; (c) tricritical density. \\

\noindent Fig.3. Phase diagrams of electrolytes  on  simple tetragonal  lattices with sublattice ordering for different degrees of anisotropy:  (a) for  $\alpha=1$ and  $\alpha=0.1$; (b) for  $\alpha=15$. Dashed lines show the metastable gas-liquid coexistence curves predicted by pure DH theory.\\

\noindent Fig.4. Phase diagrams for ionic systems on  body-centered tetragonal lattices with sublattice ordering. The gas-liquid coexistence curves predicted by pure DH theory are shown by dashed lines.

\begin{figure}[t]
\begin{center}
\vskip 1.5in
\unitlength 1in
\begin{picture}(4.5,3.3)
\resizebox{4.5in}{3.3in}{\includegraphics{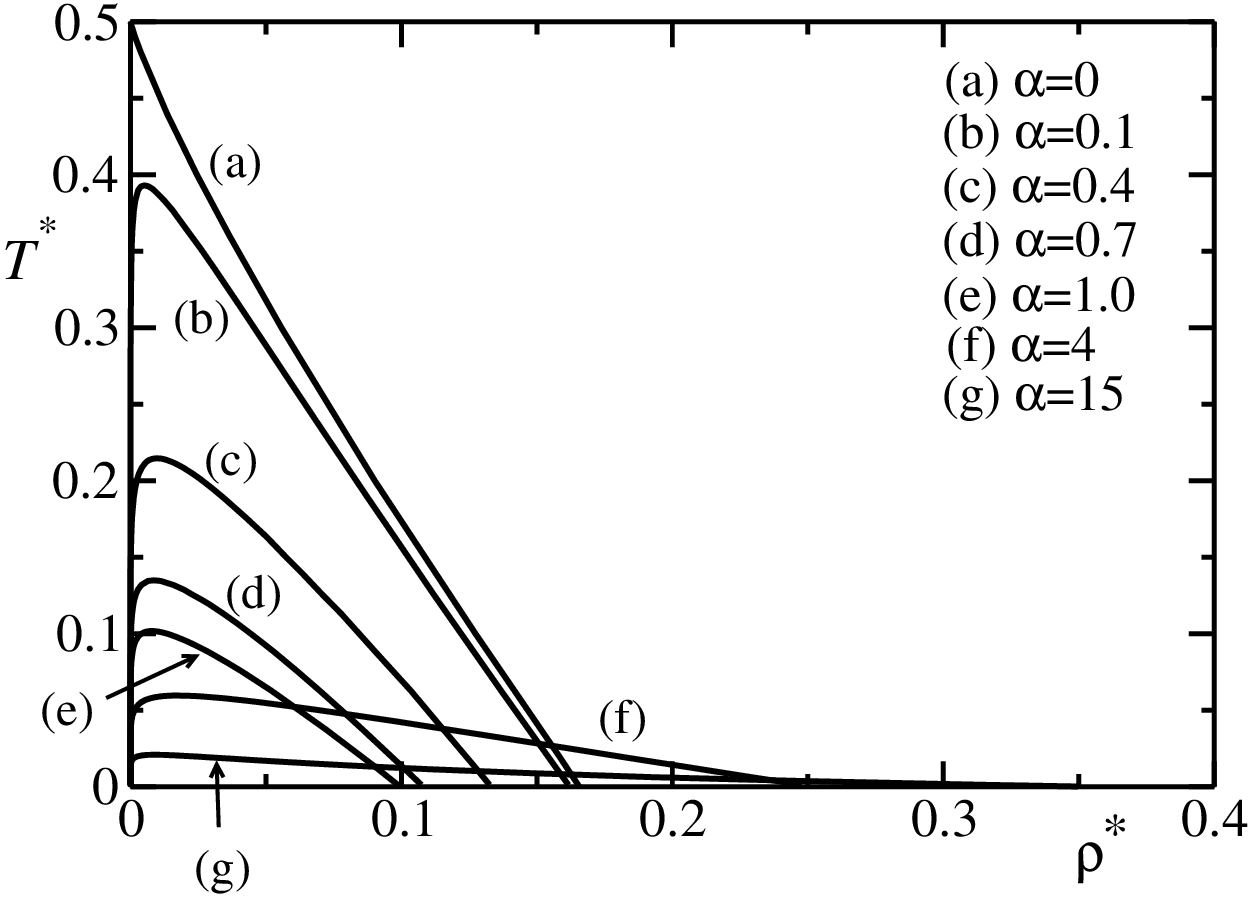}}
\end{picture}
\vskip 3in
 \begin{Large} Fig.1   \end{Large}
\end{center}
\vskip 3in
\end{figure}

\begin{figure}[t]
\begin{center}
\vskip 1.5in
\unitlength 1in
\begin{picture}(4.5,3.3)
\resizebox{4.5in}{3.3in}{\includegraphics{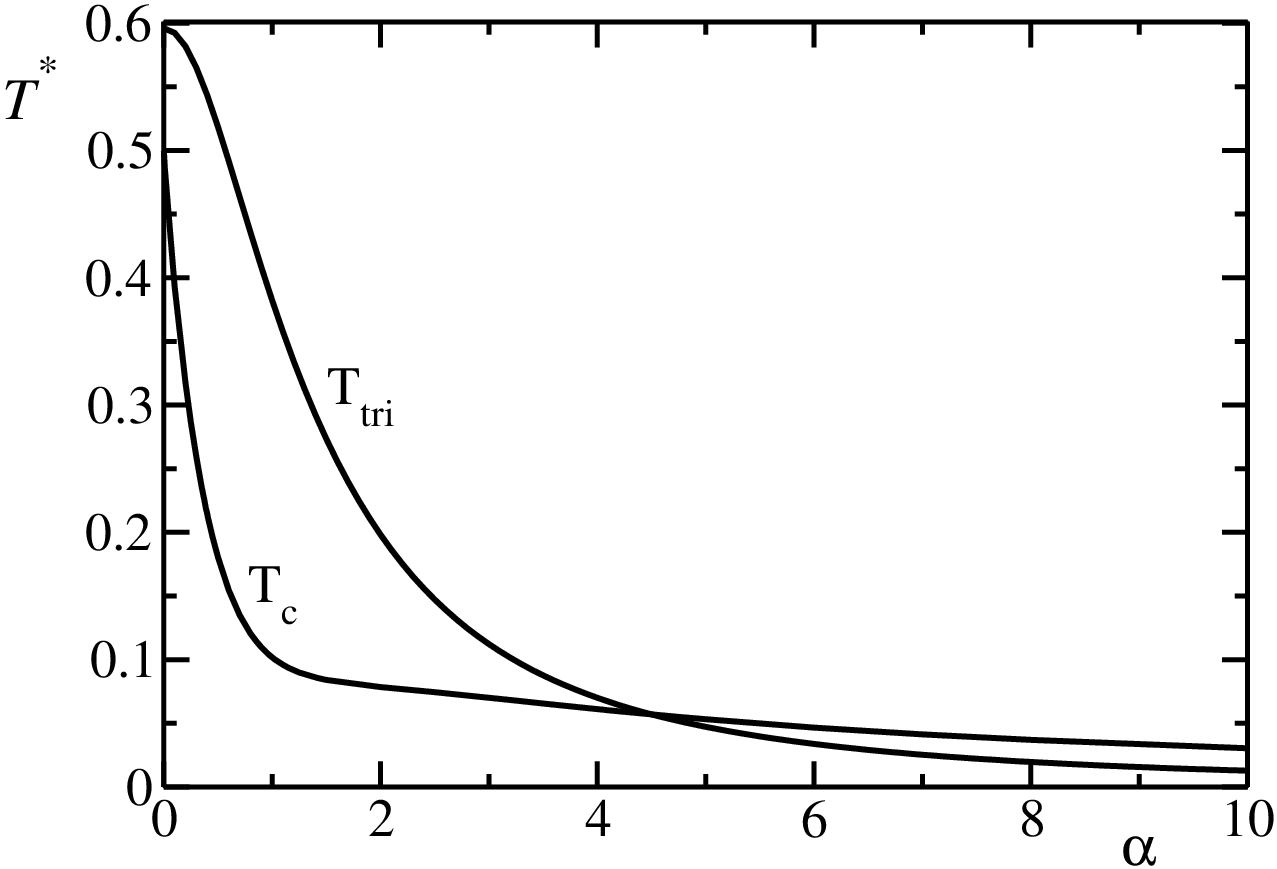}}
\end{picture}
\vskip 3in
 \begin{Large} Fig.2a   \end{Large}
\end{center}
\vskip 3in
\end{figure}

\begin{figure}[t]
\begin{center}
\vskip 1.5in
\unitlength 1in
\begin{picture}(4.5,3.3)
\resizebox{4.5in}{3.3in}{\includegraphics{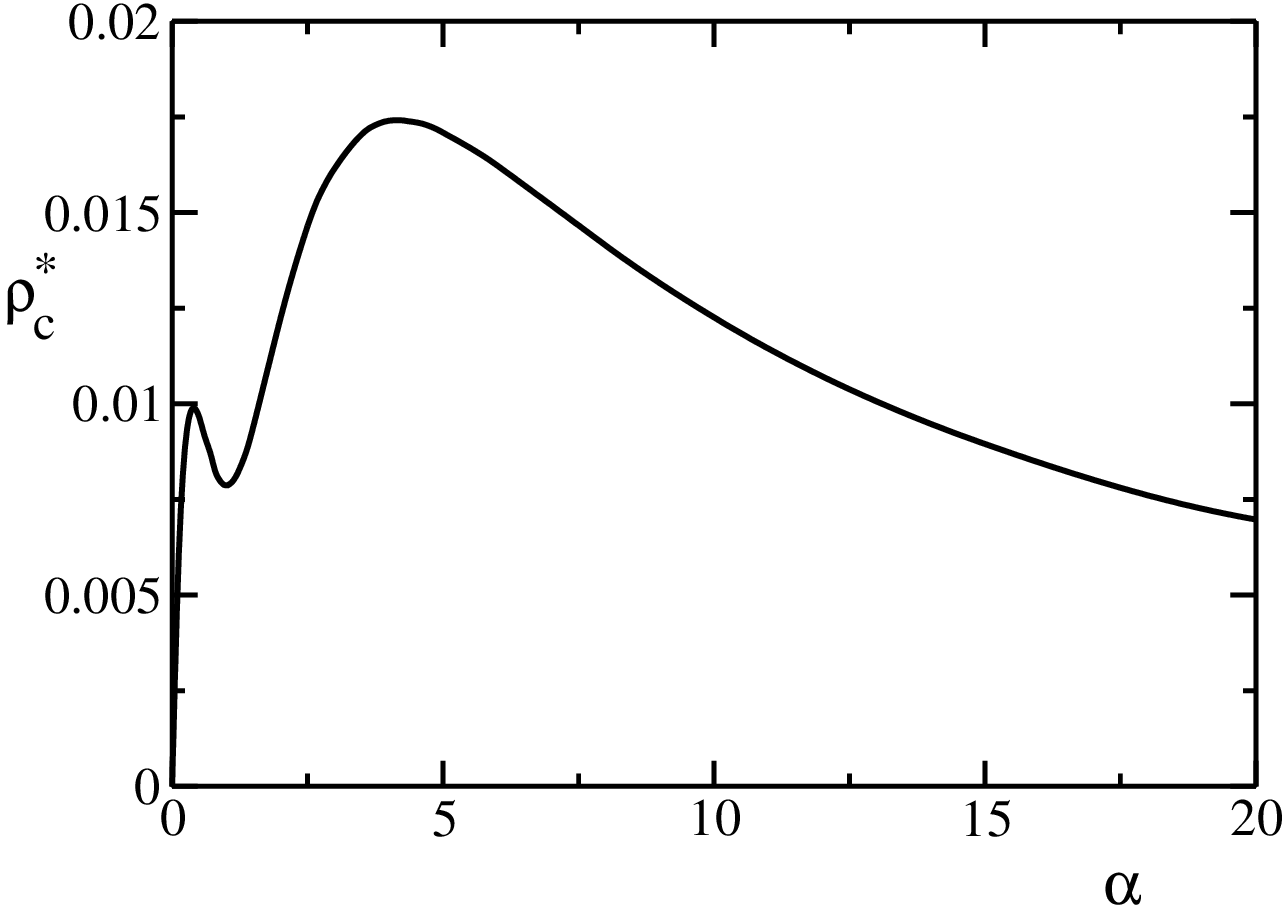}}
\end{picture}
\vskip 3in
 \begin{Large} Fig.2b  \end{Large}
\end{center}
\vskip 3in
\end{figure}

\begin{figure}[h]
\begin{center}
\vskip 1.5in
\unitlength 1in
\begin{picture}(4.5,3.3)
\resizebox{4.5in}{3.3in}{\includegraphics{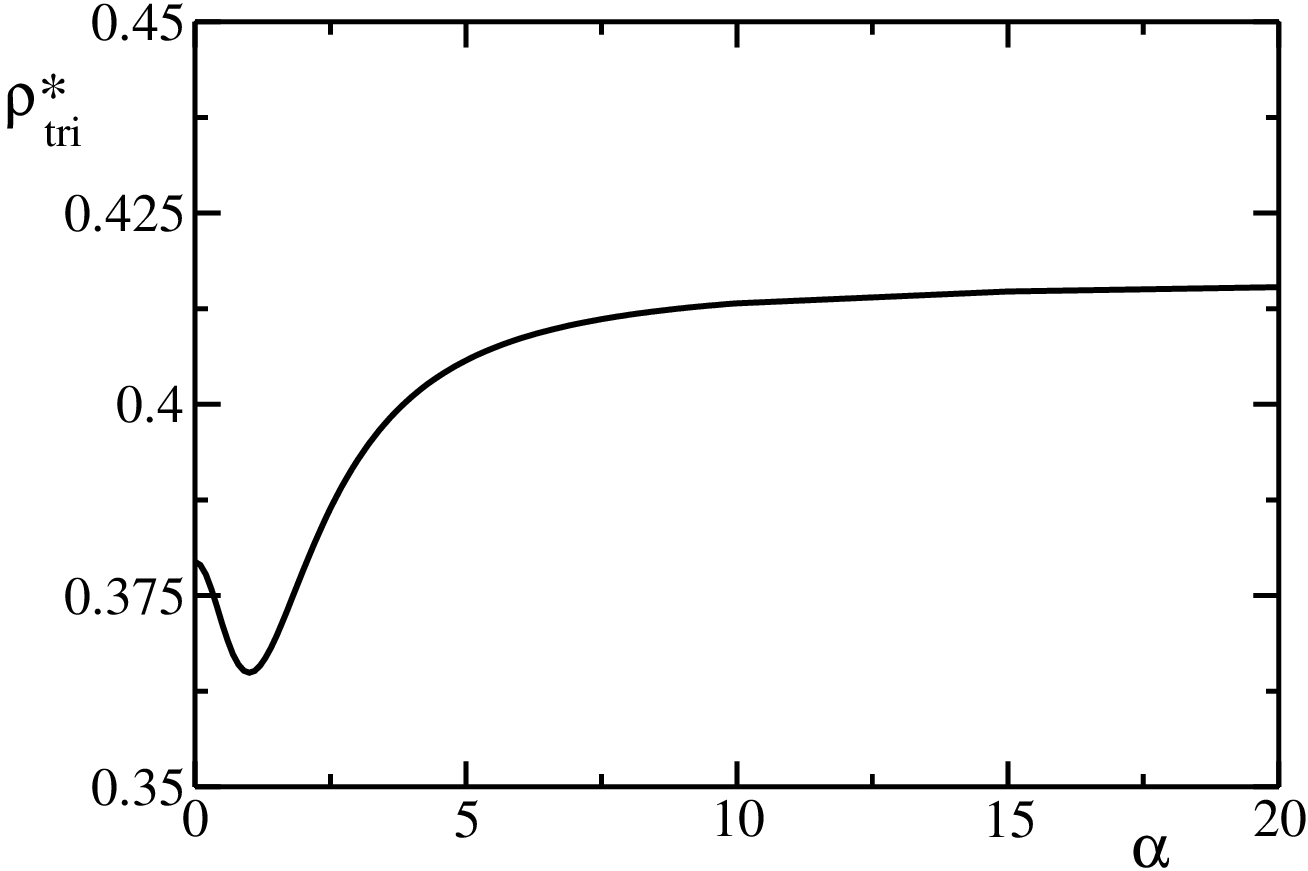}}
\end{picture}
\vskip 3in
 \begin{Large} Fig.2c  \end{Large}
\end{center}
\vskip 3in
\end{figure}

\begin{figure}[t]
\begin{center}
\vskip 1.5in
\unitlength 1in
\begin{picture}(4.5,3.3)
\resizebox{4.5in}{3.3in}{\includegraphics{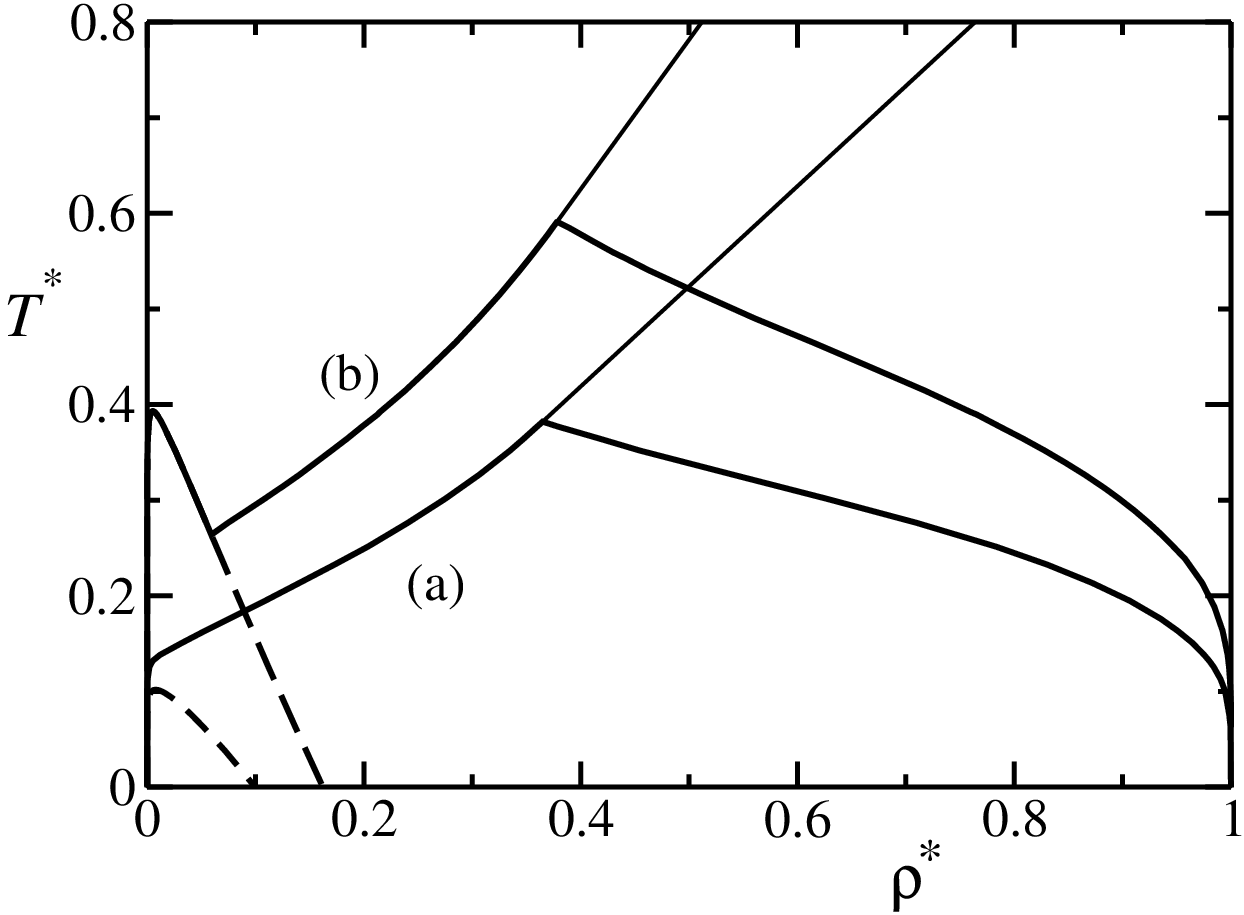}}
\end{picture}
\vskip 3in
 \begin{Large} Fig.3a  \end{Large}
\end{center}
\vskip 3in
\end{figure}

\begin{figure}[t]
\begin{center}
\vskip 1.5in
\unitlength 1in
\begin{picture}(4.5,3.3)
\resizebox{4.5in}{3.3in}{\includegraphics{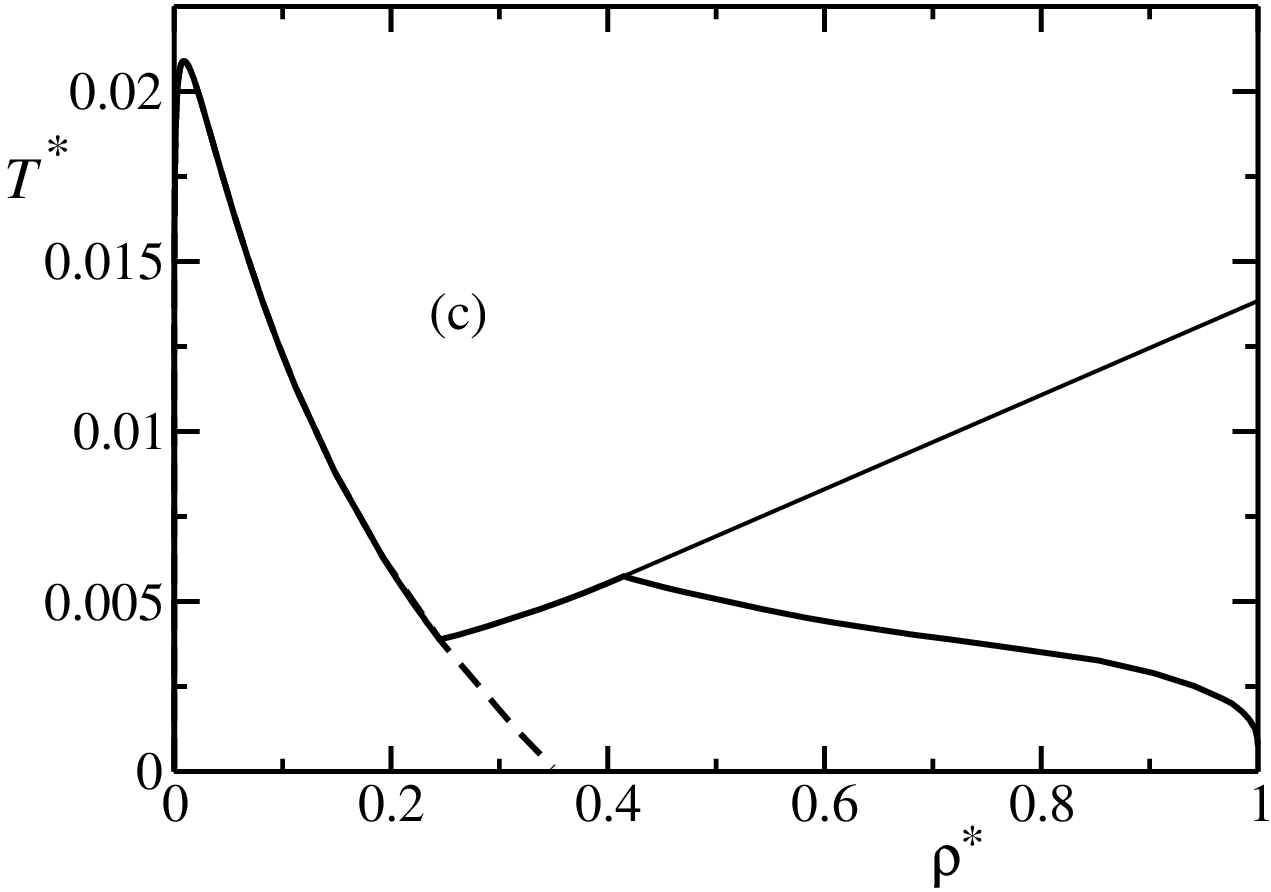}}
\end{picture}
\vskip 3in
 \begin{Large} Fig.3b  \end{Large}
\end{center}
\vskip 3in
\end{figure}

\begin{figure}[t]
\begin{center}
\vskip 1.5in
\unitlength 1in
\begin{picture}(4.5,3.3)
\resizebox{4.5in}{3.3in}{\includegraphics{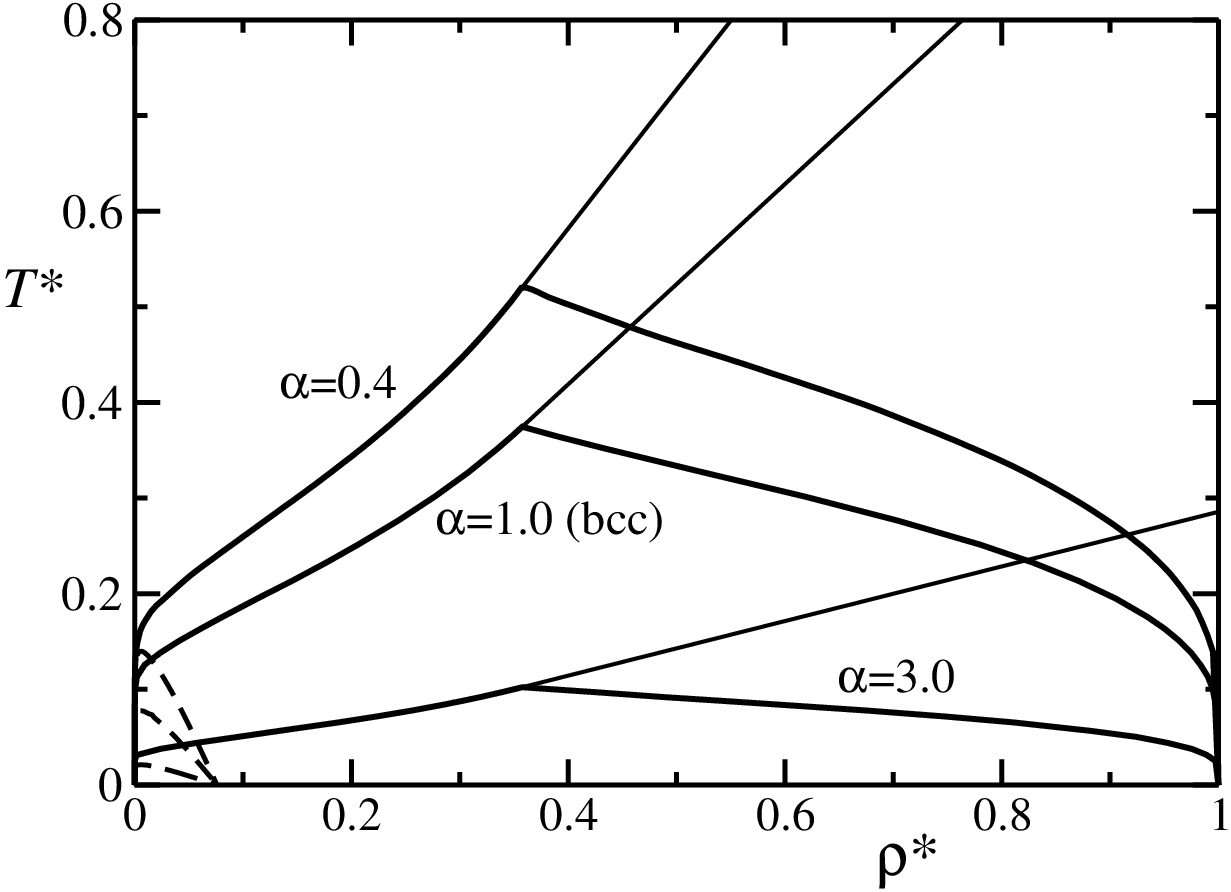}}
\end{picture}
\vskip 3in
 \begin{Large} Fig.4 \end{Large}
\end{center}
\vskip 3in
\end{figure}

\end{document}